# LearnAFE: Circuit-Algorithm Co-design Framework for Learnable Audio Analog Front-End

Jinhai Hu, *Graduate Student Member, IEEE*, Zhongyi Zhang, *Graduate Student Member, IEEE*, Cong Sheng Leow, *Graduate Student Member, IEEE*, Wang Ling Goh, *Senior Member, IEEE*, and Yuan Gao, *Member, IEEE*

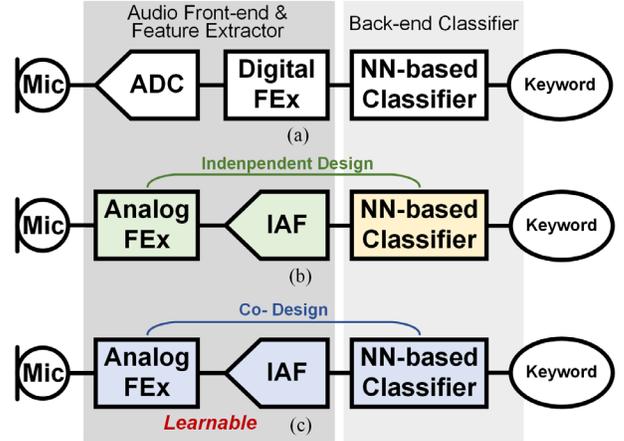

Fig. 1. (a) KWS system with ADC and digital FEx. (b) Conventional AFE designed independently from NN-based classifier. (c) Circuit-algorithm co-design for learnable AFE and NN-based classifier.

*Abstract*— This paper presents a circuit-algorithm co-design framework for learnable analog front-end (AFE) in audio signal classification. Designing AFE and backend classifiers separately is a common practice but non-ideal, as shown in this paper. Instead, this paper proposes a joint optimization of the backend classifier with the AFE's transfer function to achieve system-level optimum. More specifically, the transfer function parameters of an analog bandpass filter (BPF) bank are tuned in a signal-to-noise ratio (SNR)-aware training loop for the classifier. Using a co-design loss function $L_{BPF}$, this work shows superior optimization of both the filter bank and the classifier. Implemented in open-source SKY130 130nm CMOS process, the optimized design achieved 90.5%–94.2% accuracy for 10-keyword classification task across a wide range of input signal SNR from 5 dB to 20 dB, with only 22k classifier parameters. Compared to conventional approach, the proposed audio AFE achieves 8.7% and 12.9% reduction in power and capacitor area respectively.

## I. Introduction

AUDIO signal classification finds wide applications in areas such as human-machine interface [1], [2], environment monitoring [3], [4] and health care [5], [6]. Various types of audio signals are analyzed depending on the application, including speech, environmental sound and biomedical vital signs. Feature extraction (FEx) is a pivotal step in the classification process to identify the distinct features that represent the target signal. The classifier will be trained to identify these features with high sensitivity and specificity. In general, features are extracted either in digital or analog domain. As shown in Fig. 1(a), digital FEx requires analog-to-digital convertor (ADC) for signal acquisition, which is the bottleneck for low-power applications [7 – 10]. In Fig. 1(b), the use of analog FEx based on bandpass filterbank offers the advantage of higher energy efficiency [11 – 16] in contrast to digital FEx and is widely used for low-power edge applications. In [11 – 16], the analog front-end (AFE) for FEx and integrate-and-fire (IAF) for data conversion are designed independently from the design of the back-end classifier. Such approaches require additional non-trivial tuning and optimization at the system-level of the AFE and the classifier. Therefore, it becomes imperative to integrate both the AFE and the backend classifier in a unified manner to enable system-level optimization, as shown in Fig. 1(c).

Recently, the concept of digital learnable filter was proposed in [17 – 19]. Both time domain and frequency domain digital

Manuscript received 9 January 2025; revised 3 April 2025 and 3 May 2025; accepted 31 May 2025. This work was supported by the Agency for Science, Technology and Research (A*STAR), Singapore, partially through the Nanosystems at the Edge Program, under Grant A18A1b0055; and in part by RIE2025 Manufacturing, Trade and Connectivity (MTC) Programmatic Fund, High Linearity Silicon Germanium Photonic Modulator for 6G Analog Radio over Fiber Project, under Grant M24M8b0004. (Corresponding Author: Yuan Gao)

J. Hu is with the School of Electrical and Electronic Engineering, Nanyang Technological University, Singapore 639798 and he is also with the Institute of Microelectronics (IME), Agency for Science, Technology and Research (A*STAR), 2 Fusionopolis Way, Innovis #08-02, Singapore 138634, (e-mail: jinhai001@e.ntu.edu.sg/ hujh@ime.a-star.edu.sg).

Z. Zhang is with the School of Electrical and Electronic Engineering, Nanyang Technological University, Singapore 639798 and he is also with the Institute of Microelectronics (IME), Agency for Science, Technology and Research (A*STAR), 2 Fusionopolis Way, Innovis #08-02, Singapore 138634, (e-mail: zhongyi001@e.ntu.edu.sg).

C. S. Leow was with the Institute of Microelectronics (IME), Agency for Science, Technology and Research (A*STAR), 2 Fusionopolis Way, Innovis #08-02, Singapore 138634. He is now with the Department of Electrical and Computer Engineering, University of Michigan, Ann Arbor, MI 48109 USA (email: leowcs@umich.edu)

W. L. Goh is with the School of Electrical and Electronic Engineering, Nanyang Technological University, Singapore 639798 (e-mail: ewlgoh@ntu.edu.sg).

Y. Gao is with the Institute of Microelectronics (IME), Agency for Science, Technology and Research (A*STAR), 2 Fusionopolis Way, Innovis #08-02, Singapore 138634, (e-mail: gaoy@ime.a-star.edu.sg).

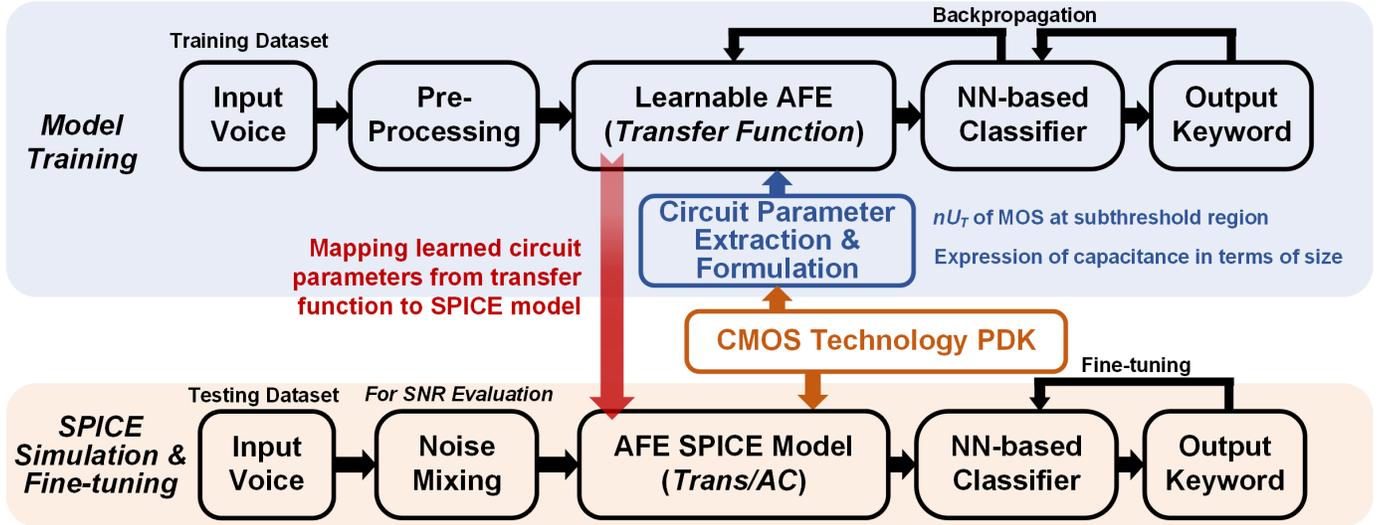

Fig. 2. Block diagrams of the learnable audio analog front-end design framework.

filterbanks were optimized through representation learning using backpropagation in deep neural networks [20]. The learned filters reduced the effect of over-fitting and obtained better performance than fixed filterbank [21].

Inspired by the success of learnable digital filter banks, this work introduces a circuit-algorithm co-design framework for learnable audio analog front-end for low-power applications. The parameters of a differential super source-follower bandpass filter (DSSF-BPF) bank are optimized together with the neural network classifier in a signal-to-noise ratio (SNR)-aware training process for optimal system performance. This paper is an expanded version of [22] with additional analysis and significant improvements on design formulation, training strategy and circuit simulation. The contributions of this work can be summarized as follows:

- This paper introduces a novel co-design framework combining analog filterbank design with a lightweight neural network classifier. This approach optimizes the analog filterbank parameters in tandem with the classifier, achieving a system-level performance improvement rather than optimizing components independently.
- A new co-design loss function, $L_{BPF}$, is proposed to integrate classification loss, power loss, and area loss into the training process. This co-optimization ensures that the analog front-end is efficient in power and area usage while maintaining high classification accuracy under various SNRs.
- This work employs Bayesian optimization (BO) for hyperparameter tuning, focusing on learning rate, weight decay, and regularization coefficients for the loss components. This approach balances hardware performance and classification accuracy efficiently. Based on 25 trials, the process enhances classification accuracy from 87.4% to 94.2% in keyword spotting (KWS) applications.

Moreover, all circuit parameters are extracted from SKY130 PDK [23]. The learned AFE's netlist and performance are verified on Ngspice based on AC and transient responses [24]. To the best of our knowledge, this is the first reported learnable analog front-end using circuit-algorithm co-design strategy.

## II. PROPOSED FRAMEWORK

Fig. 2 shows the proposed closed-loop circuit-software co-design framework for learnable AFE design. This design framework consists of two stages. The first stage is to train the transfer function of AFE together with the neural network classifier. The trained transfer function parameters are mapped to the actual circuit SPICE model parameters. The second stage performs transistor level circuit simulation and fine tune the neural network parameters. Google Speech Command Dataset [25] (GSCD) which has 16 kHz sampling rate with 16-bit resolution is used in the training process of both stages. In the first stage, the input training dataset is firstly pre-processed with resampling, data augmentation and SNR tuning (explained further in Section III). The data is then passed through learnable AFE that is formulated from the CMOS technology PDK. The neural network (NN)-based classifier is trained using these filtered data for KWS with the knowledge of the AFE's transfer function. By incorporating the parameters of the AFE in the training process, the AFE is optimized with the KWS results during each training iteration. During the second stage, the SPICE model of the AFE with the trained parameters is used to generate the output features before passing through the classifier. Noise can be mixed at original waveforms to test the robustness of the model through SNR evaluation. The performance degradation caused by transitioning from an ideal transfer function to a SPICE model is compensated by fine-tuning the classifier.

Fig. 3 shows the detailed circuit architecture of the proposed learnable audio AFE, back-end classifier and the schematic of the differential super source-follower bandpass filter (DSSF-BPF). Each FEx channel comprises a tunable DSSF-BPF and a spectrogram generator. The DSSF-BPF is strategically adopted for its performance, tunability, and efficiency [12], [26]. The generated spectrogram is obtained using a half-wave rectifier (HWR) and integrate-and-fire (IAF) layer before the classifier

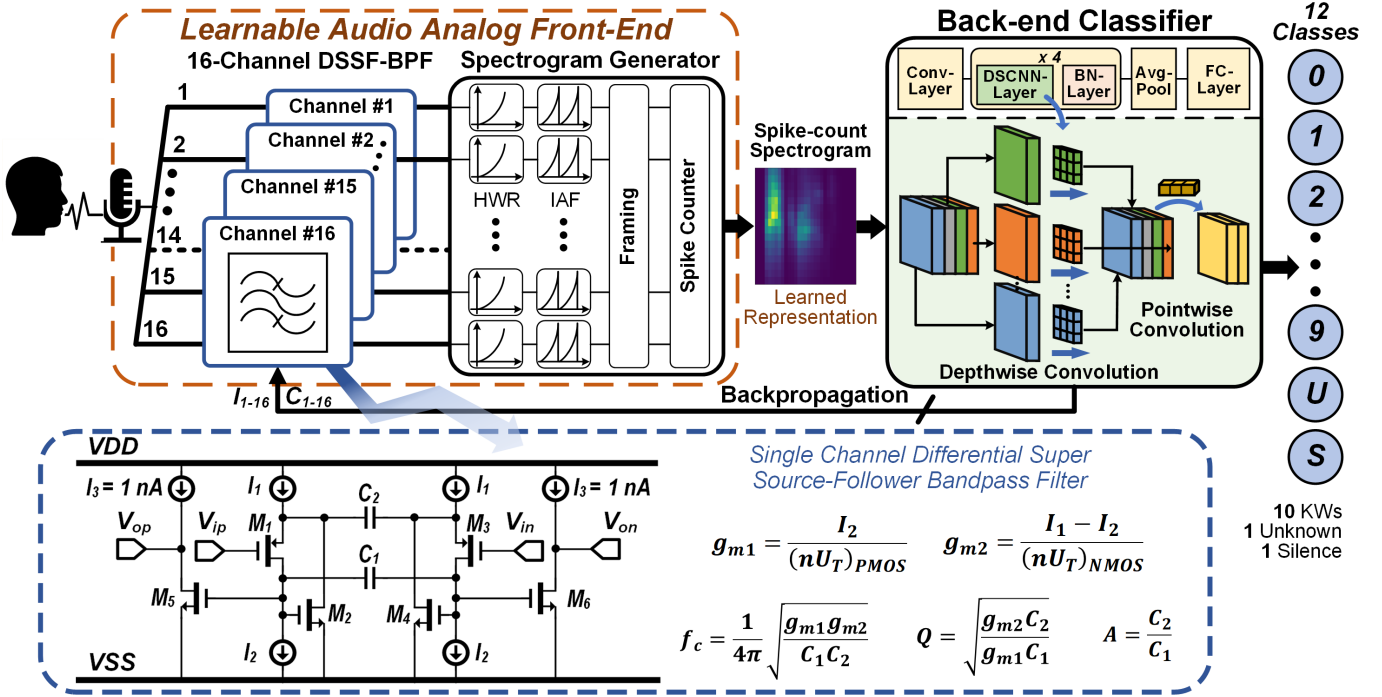

Fig. 3. The block diagram of the proposed learnable audio analog front-end (LearnAFE) and the schematic of the differential super source-follower bandpass filter (DSSF-BPF).

[11], [13]. The HWR is modelled as a ReLU function, and the IAF layer is approximated by summing the rectified outputs over fixed length of frames and converting to the spike count, thereby maintaining end-to-end differentiability during training. A depthwise separable convolutional neural network (DSCNN) is used for 12-classes (10 keywords + 1 unknown + 1 silence) classification.

The DSCNN model encompasses one convolutional layer (Conv-Layer), four DSCNN blocks, followed by one averaging and pooling layer (Avg-Pool) and one fully connected layer (FC-Layer). Each DSCNN block contains one DSCNN layer and one batch normalization layer (BN-Layer). The schematic of the 2nd-order DSSF-BPF is shown in the bottom of Fig. 3. Transistors $M_{1,2}$ and $M_{3,4}$ serve as the super-source-followers, and the floating capacitors $C_{1,2}$ determine the pole locations. The transistor transconductance for $M_{1,2}$ can be expressed in $g_{m1}$ and $g_{m2}$ as shown in Fig. 3, where $n$ is the subthreshold slope factor, and $U_T$ is thermal voltage at room temperature.

Through small signal model analysis, and neglecting body effects, the filter transfer function $H(s)$ is expressed in (1). Likewise, the central frequency ($f_c$), the quality factor ($Q$) and the passband gain ($A$) can be derived as shown in Fig. 3 (bottom right). (1) reveals that all parameters ($g_{m1}$, $g_{m2}$, $C_1$, $C_2$) within this structure can be adjusted individually. Operating in the subthreshold region, the $g_m$ of the DSSF-BPF can be tuned through external current adjustments.

$$H(s) = -\frac{\frac{g_{m1}}{2C_1}s}{s^2 + \frac{g_{m1}}{2C_2}s + \frac{g_{m1}g_{m2}}{4C_1C_2}} \quad (1)$$

It can be observed that the key performance of each DSSF-BPF can be largely defined by four main circuit parameters ($g_{m1}$, $g_{m2}$, $C_1$, $C_2$). These parameters are included in the system training process using backpropagation. With the initial parameters based on the reference design in SPICE circuit simulation, the parameters and resulting signal output are updated during the training process. This joint optimization of the AFE and the classifier aims to reduce incompatibility between the two components when optimized separately.

III. METHODOLOGY

A. Data Preprocessing

The Google Speech Command Dataset (GSCD) [25] is used in the training and verification process. Three pre-processing techniques are employed for speech comments during the training phase. First, the data is resampled from 16 kHz to 20 kHz to match with the BPF SPICE simulation output which is sampled at the Nyquist rate (20 kHz) of the BPF upper frequency limit (10 kHz).

Following this, each waveform is augmented by padding with 0.1-second blank segments at both the beginning and end [27]. A 1-second random crop window is applied during training to enhance robustness on time shifting of commands and reduce overfitting of model on training dataset.

Finally, the training data is augmented by incorporating background noise to improve noise robustness. The GSCD offers six 1-minute background noise audio files, which are randomly segmented and mixed into the original waveform at varying SNRs. This approach allows us to train the model to be robust to noise across a range of SNR levels.

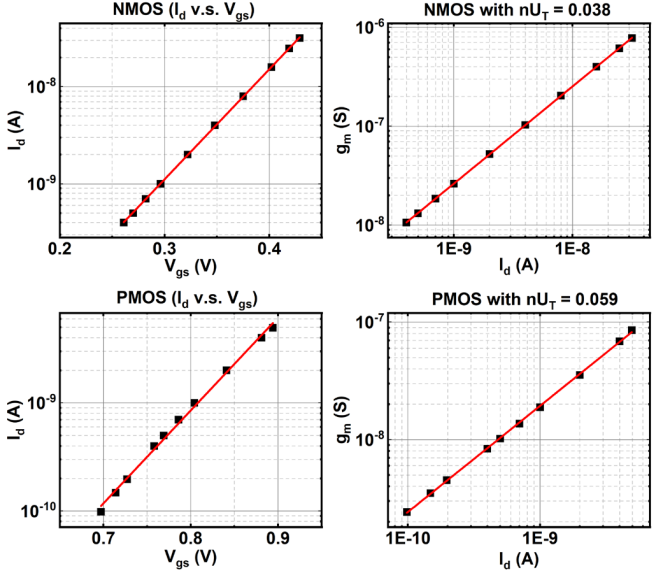

Fig. 4. NMOS and PMOS behaviour at subthreshold region simulated from PDK and curve fitting results.

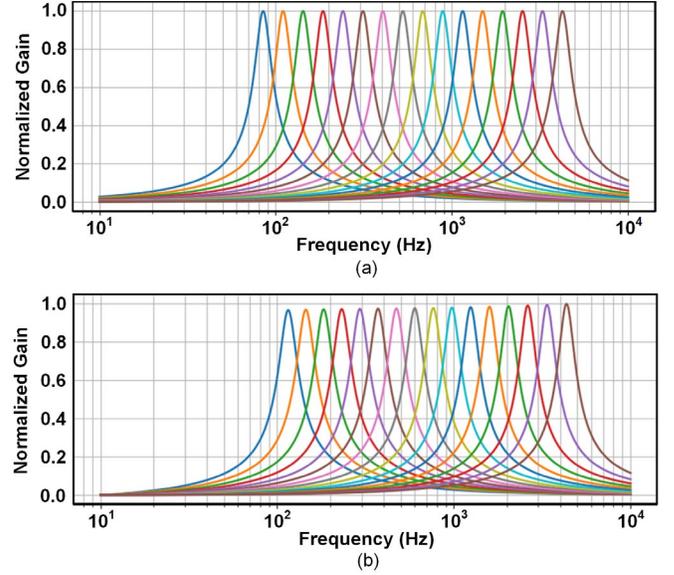

Fig. 6. AC response of 16 channel DSSF-BPF generated from (a) ideal transfer function and (b) SPICE simulation on same initial circuit parameters.

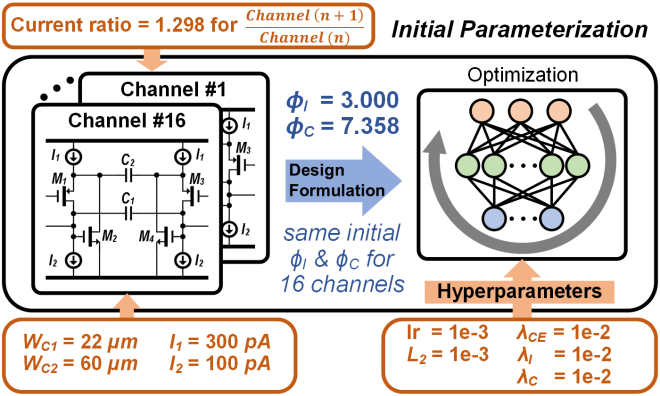

Fig. 5. Initial parameterization for AFE circuit design and hyperparameters in training optimization.

At the classification level, a category labelled as "unknown" is assigned to all speech commands except for keywords, while a category called "silence" is introduced to capture samples from the background noise. Consequently, the system is tasked with classification across a total of 12 classes, encompassing ten keywords, one "unknown" category, and one "silence" category, as illustrated in Fig. 3.

*B. AFE Design Formulation*

For each AFE channel, $g_{m1}$, $g_{m2}$, $C_1$, $C_2$ can be formulated using the open-source SKY130 PDK [23]. As an improvement from our earlier work [22], this work considers the actual hardware implementation much more closely. Instead of using transconductance ($g_{m1}$ and $g_{m2}$) in the earilier work, BPF current ($I_1$ and $I_2$) is used to allow for actual tuning through external currents. Similarly, the capacitor sizing ($W_{C1}$ and $W_{C2}$) is used instead of capacitance ($C_1$ and $C_2$) to consider the nonideality of MIM capacitors. The transistor transconductance ($g_m$) is the first order deviation of drain current ($I_d$) with respect to Gate-Source Voltage ($V_{gs}$), as shown in (2) and (3). Thus, the relationship between $g_{m1,2}$ and $I_{1,2}$ is expressed in Fig. 3. $(nU_T)_{NMOS} = 0.038$ V and $(nU_T)_{PMOS} = 0.059$ V are obtained based on regression analysis of the SPICE simulation of transistors working in subthreshold region, as shown in Fig. 4.

$$I_d = I_{d0}\frac{W}{L}\exp\left(\frac{V_{gs}-V_{th}}{nU_T}\right) \quad (2)$$

$$g_m = \frac{\partial I_d}{\partial V_{gs}} = \frac{I_d}{nU_T} \quad (3)$$

The capacitance $C$ can be expressed in (4) in terms the capacitor's physical size, $W_C$ and $L_C$.

$$C = [(W_C \times L_C) \times t + (W_C + L_C) \times C_{eff}] \times 10^{-3} \quad (4)$$

where $t$ is the thickness and $C_{eff}$ is the coefficient of fringing effect. The two constant coefficients can be extraced from CMOS technology PDK. In this work, every capacitor is assumed to have a square shape, such that $W_C = L_C$, for simplicity in both optimizaiton and actual layout consideration.

The optimization boundaries such as $f_c$ spans from 100 Hz to 5k Hz. It should be noted that some filter parameters such as $I_{1,2}$ and $W_{C1,2}$ may have huge value differences up to three order or even more. Such big differences complicate training due to the impact on the learning rates and potentially leading to vanishing/exploding gradients [28]. To address this issue, two trainable scaling factors ($\phi_I$ and $\phi_C$) are introduced in (5), which are defined as the ratio to the baselines, $I_2$ and $C_1$, for each channel.

$$\phi_I = \frac{I_1}{I_2}, \quad \phi_C = \frac{C_2}{C_1} = \frac{W_{C2}^2 \times t + W_{C2} \times 2C_{eff}}{W_{C1}^2 \times t + W_{C1} \times 2C_{eff}} \quad (5)$$

Both factors are scalable in real circuit design. This approach enables a flexible and realistic representation of current values being imposed during training ($\phi_I$>1) such that the optimization result adhere to circuit theory. $\phi_C > 1$ is applied as a constraint for optimization to ensure the capacitor size of $C_2$ is not unrealistically small during optimization. Moreover, while learning circuit values involves twice as many parameters for 16 channels (64 parameters) compared to scaling factors (32 parameters), typical circuit values tend to have larger magnitudes than neural network parameters, posing challenges for constraining and gradient vanish during training. Therefore, reducing power and area is more productive through learning using the scaling factors.

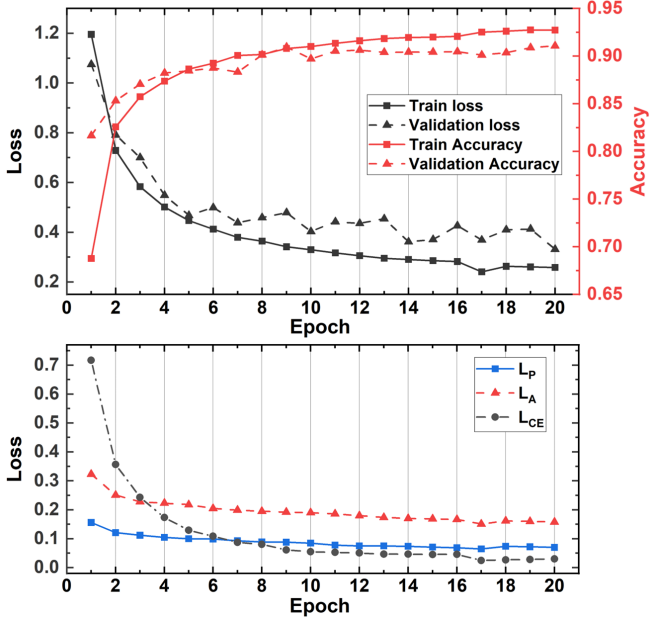

Fig. 7. (top) Training and validation performance over 20 epochs. (bottom) Loss reduction of $L_P$, $L_A$ and $L_{CE}$ through training.

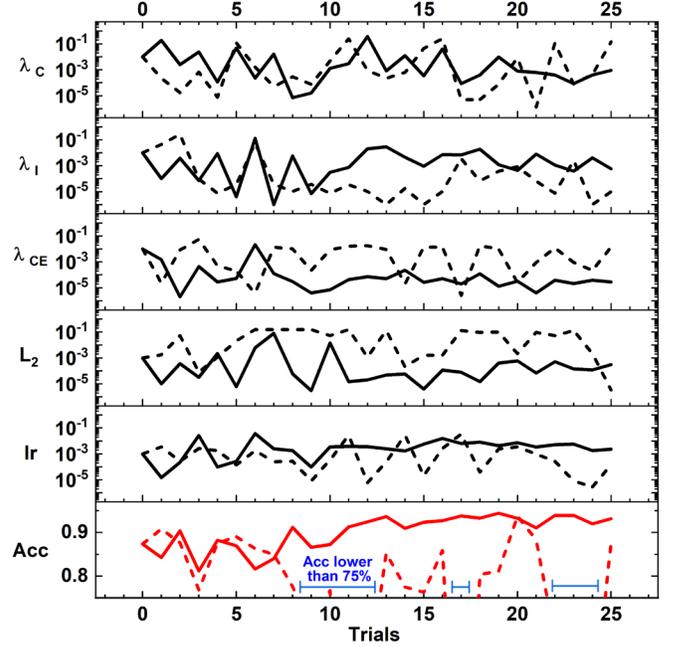

Fig. 8. Hyperparameter tuning process through 25 trials, with solid and dashed lines representing BO and random search, respectively.

The corresponding $g_{m2}$ used in transfer function (1) can be expressed in terms of $g_{m1}$, $\phi_I$, $(nU_T)_{NMOS}$ and $(nU_T)_{PMOS}$, as listed in (6). As $g_m$ of a transistor is difficult to tune directly when the hardware is fabricated, we use the current $I_{1,2}$ to change the filter's performance as they can be controlled with external current sources. With each channel's $\phi_I$ and $\phi_C$, the DSSF-BPF characteristics can be written as (7).

$$g_{m2} = (\phi_I - 1) \times g_{m1} \frac{(nU_T)_{PMOS}}{(nU_T)_{NMOS}} \quad (6)$$

$$f_c = \frac{1}{4\pi} \frac{I_2}{C_1} \sqrt{\frac{\phi_I - 1}{(nU_T)_{NMOS} \cdot (nU_T)_{PMOS} \cdot \phi_C}}$$

$$Q = \sqrt{\frac{(nU_T)_{PMOS}}{(nU_T)_{NMOS}} \cdot \phi_C \cdot (\phi_I - 1)}, \quad A = \phi_C \quad (7)$$

The initial parameters for DSSF-BPF configuration are shown in Fig. 5, including the initial value for circuit parameters and scaling factors. The current ratio between consecutive channels is 1.298, to ensure that the central frequencies of 16-channel DSSF-BPF vary from 100 Hz to 5k Hz. The scaling factors represent the ratio of two coupled parameters, allowing for more balanced updates during the training process. The frequency response with ideal transfer function involving these initial values is shown in Fig. 6(a), which showcases uniform $A$ and $Q$ across 16 channels.

### C. Co-designed Loss Function

To concurrently optimize the classifier and the AFE performance, a novel loss function, $L_{BPF}$, is proposed. In addition to the cross-entropy loss ($L_{CE}$) for classifier, the AFE power loss ($L_P$) and area loss ($L_A$) are incorporated into $L_{BPF}$ to represent the hardware efficiency. Specifically, as depicted in (8), $L_P$ and $L_A$ are formulated using the scaling factors $\phi_I$ and $\phi_C$, which are proxies for current ratios and capacitance values in the circuit. The power consumption of DSSF-BPF, expressed as $2V_{DD}(I_1 + I_3)$, is directly proportional to $\phi_I$. Similarly, $\phi_C$ encapsulates the area contributions of capacitors, which form a significant portion of the circuit layout. To balance the influence of these terms, regularization coefficients $\lambda_{CE}$, $\lambda_I$, $\lambda_C$ are introduced to adjust the importance of each loss components during model initialization.

$$L_{BPF} = L_{CE} + L_P + L_A$$
$$\rightarrow L_{BPF} = \lambda_{CE} L_{CE} + \lambda_I \sum_{i=1}^{16} \phi_{I,i} + \lambda_C \sum_{i=1}^{16} \phi_{C,i} \quad (8)$$

Fig. 7 illustrates the trends in accuracy and loss for both the training and validation datasets over 20 epochs. With a fixed set of regularization coefficients, the $L_{BPF}$ is progressively minimized in training, resulting in a simultaneous reduction across all three loss components. A lower cross-entropy loss corresponds to improved classification accuracy, a reduced power loss translates to lower AFE power consumption, and a decreased area loss reflects a smaller capacitor footprint in the 16-channel BPF bank. Since $\phi_I$ and $\phi_C$ are defined as the ratio to baseline parameters ($I_2$ and $C_1$), minimizing these terms during training directly lowers total currents and capacitances, thereby reducing both power consumption and chip area in the final circuit.

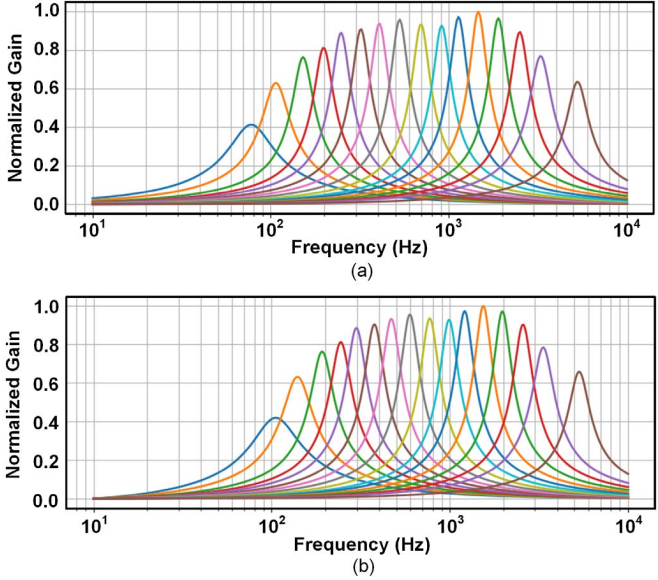

Fig. 9. AC response of 16 channel DSSF-BPF generated from (a) ideal transfer function and (b) SPICE simulation on same learned circuit parameters.

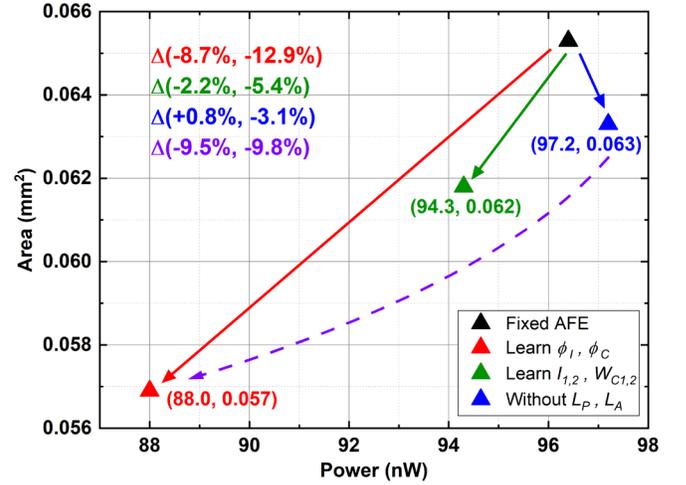

Fig. 10. Comparison of hardware utilization based on different learning strategies.

The comprehensive formulation incorporates the equivalent circuit transfer function, adaptive scaling strategies, and system-level optimization, offering an integrated perspective of the interaction between the analog filter's characteristics and the machine learning framework.

*D. Hyperparameter Tuning*

There are five hyperparameters in co-design training, including learning rate ($lr$), weight decay ($L_2$), and regularization coefficients for three loss components ($\lambda_{CE}$, $\lambda_I$, $\lambda_C$). The complete loss function used during backpropagation includes an $L_1$ regularization term ($L_{backward} = L_{BPF} + L1$). Thus, independently tuning these hyperparameters helps us achieve a better overall performance that jointly satisfies accuracy, hardware constraints, and complexity considerations. Proper hyperparameter tuning ensures that the model achieves a balance between performance metrics and resource efficiency. Parameter tuning is often done with straightforward methods such as grid search [29]. However, grid search faces difficulty when the dimensionality of the parameter space increases, particularly when dealing with continuous parameters. On the other hand, Bayesian optimization (BO) enables parameter tuning in relatively few iterations by constructing a smooth model from an initial set of parameterizations [30]. This smooth model decides what parameterizations to evaluate next based on observations from previous evaluations. As contrast to random search, it serves as a more explainable approach to justify design choices.

In this work, we adopt an open-source Python algorithm library Ax [31], [32] to implement the BO for hyperparameter tuning. The tuning process is illustrated in Fig. 8 (solid lines). Trial 0 denotes the performance with initial parameters given in Fig. 5. Next, BO constructs a smooth model of outcomes using Gaussian processes based on the observations from next 10 rounds of trials, even if it's noisy, such as trial 3 and trial 6. The smooth model enables predictions at unobserved parameterizations and quantifies uncertainty around them. These predictions and uncertainty estimate feed into an acquisition function, which evaluates the value of observing a specific parameterization for the next 15 trials. Expected improvement (*EI*) is chosen as the acquisition function in this work. It incentivizes the evaluation of the objective function (*f*) based on the expected improvement relative to the current best. If $f^*$ represents the current best observed outcome, and our objective is the maximize *f*, then *EI* is defined as:

$$EI(x) = E[\max(Acc(x) - Acc^*, 0)] \qquad (9)$$

where $x$ is a potential parameterization, classification accuracy (*Acc*) represents objective function *f*, *Acc(x)* is the predicted outcome for that parameterization, and *E[.]* denotes the expectation. The parameterization with the highest *EI* is selected and evaluated in the next step. Once the parameter space is adequately explored, *EI* narrows in on locations where the objective function is converged. As shown in Fig. 8, comparing to the set of initialized hyperparameters in trial 0, the accuracy is enhanced from 87.4% to 94.2% based on 25 tuning trials.

We repeated the BO procedure across multiple random seeds and observed negligible variations in the final hyperparameter configurations. These results suggest that the optimization process is stable and robust against the effects of random initialization. Moreover, we conducted a benchmarking comparison between random search and BO, as shown in Fig. 8 (dash lines), demonstrating that BO achieved better validation accuracy and converged to a higher value within the same 25 trials. Thus, BO represents a pragmatic choice in terms of computational efficiency and optimization performance.

*E. SPICE Verification and Finetuning*

To evaluate the AFE performance more accurately including circuit non-idealities, transistor-level SPICE simulation is performed during inference phrase as shown in Fig. 2. The optimized circuit parameters are written into a spice file during the initialization of the AFE SPICE model, incorporating AFE circuit parameters as external inputs. A separate spice file is generated with parameters for the circuit of 16 channels, each channel comprising $I_{1,2}$ and $W_{C1,2}$. AC response and transient simulation are performed for a 16-channel DSSF-BPF using Ngspice [24]. The output of the AC response simulation is the transfer function of DSSF-BPF. With the same initial circuit parameters listed in Fig. 5, the initial AC response in Fig. 6(b) reveals slightly unflatten gain across the 16 channels compared to Fig. 6(a). During inference, the nonideal transfer function replaces the ideal transfer function as FEx, followed by the trained DSCNN as a back-end classifier. The AC response of AFE with the learned circuit parameters is shown in Fig. 9(b).

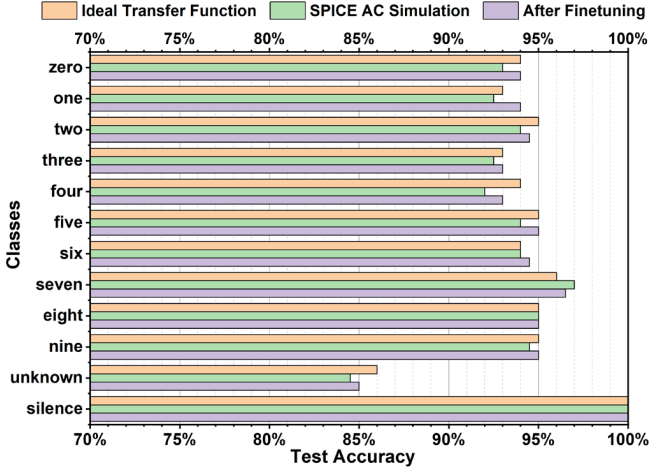

Fig. 11. KWS accuracy using learned AFE through ideal transfer function, SPICE AC simulation, and finetuned model.

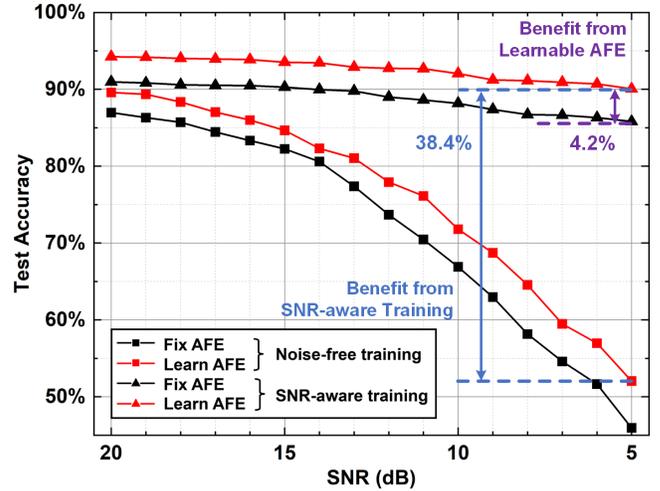

Fig. 12. KWS accuracy obtained over different SNR levels.

During the transient simulation, the AFE SPICE model accepts the speech command waveform as a voltage input. Sixteen filtered waveforms are generated through simulation with a 50 μs linearized time step and passed through a spectrogram generator as depicted in Fig. 3. Subsequently, the spike-count feature map passes through the same trained DSCNN to obtain the classification result.

Both Fig. 6 and Fig. 9 show differences between the frequency response of the transfer function and the SPICE simulation results. These discrepancies caused a drop in the classification accuracy by approximately 1% using the same trained back-end classifier. To restore the accuracy, finetuning was performed by adjusting only the back-end classifier for 5 epochs using the training dataset. The input waveforms were converted into spectrograms using the learned AFE through SPICE simulation. The model optimizer observed that gradients were computed exclusively for the parameters of the DSCNN classifier. Additionally, the loss function considered only cross-entropy loss, as the AFE remained fixed during fine-tuning.

IV. RESULTS AND DISCUSSION

*A. Co-Design Results using Transfer Function*

Fig. 6(a) and Fig. 9(a) illustrate the transfer function of 16-channel DSSF-BPF before and after training. Initially, $\phi_I$, $\phi_C$, and $Q$ are the same across 16 channels (Fig. 6(a)), while learned AFE shows non-uniform gains and Q-factor across 16 channels (Fig. 9(a)). The learned DSSF-BPF exhibits nonuniform gain, with reduced amplification in the lower and higher frequency channels compared to the mid-frequency channels. Notably, a significant roll-off occurs around 1 kHz, aligning with the understanding that lower frequencies often correspond to background noise and may carry less distinctive information pertinent to keyword recognition. This characteristic of the BPF is achieved by leveraging diverse representations across the 12 classes. During training, the co-design algorithm reallocates emphasis to the more informative frequency bands. While all samples from each keyword exhibit similar represented features, the "unknown" category encompasses all features from the remaining 25 keywords. Likewise, the "silence" category contains various segments from six background noise samples. The AFE is optimized to tolerate differences within the "unknown" class and the "silence" class, but distinguishes the representations from the ten specific keywords.

Fig. 10 shows the hardware utilization using different learning strategies. Compared to training the circuit component values ($I_{1,2}$ and $W_{C1,2}$) directly in [22], the learning of scaling factors ($\phi_I$ and $\phi_C$) achieves notable reductions of 8.7% and 12.9% in DSSF-BPF power and area consumption, respectively. Similarly, the use of a system-level loss function contributes significantly to the power and area optimization within the LearnAFE framework. As high-frequency channels consume more power due to their large initial bias current, optimizing these channels can notably reduce the AFE's power consumption. A naïve approach of training the AFE along with the KWS classifier led to an increase in power consumption by 0.8% with a 3.1% reduction in capacitor area. Compared to training solely with cross-entropy loss (blue symbol in Fig. 10), incorporating a system-level loss function can decrease the total power consumption and capacitor area by 9.5% and 9.8%, respectively, as shown in the violet line in Fig. 10.

*B. SPICE Simulation Results*

The AC response of the 16-channel DSSF-BPF generated from SPICE simulation on the same learned circuit parameters is shown in Fig. 9(b). The discrepancy between the transfer function and AC response primarily manifests as a minor frequency shift and gain difference, particularly noticeable in the low-frequency range. Despite the slight difference between them, the majority of energy captured by the same filter channel corresponds to the same frequency range. Additionally, analog BPFs typically exhibit a wide bandwidth and non-zero gain across the entire frequency range. The minor shift in the central frequency also results in a slight variation in channel-wise energy. By finetuning the classifier as discussed in Section III.E, the KWS accuracy improved to 94.2% at 20 dB SNR and 90.5% at 5 dB SNR. Consequently, the drop in classification accuracy is less than 0.2% when using the SPICE model of learned audio AFE. The performance of KWS among the ideal transfer function, SPICE simulation, and finetuned model is compared in Fig. 11.

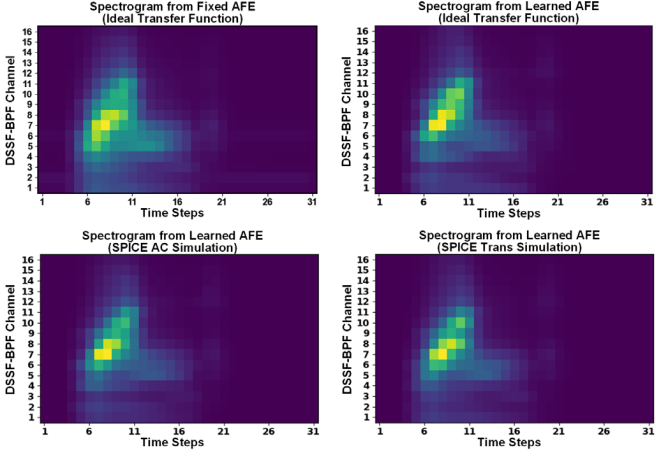

Fig. 13. Spike-count spectrogram of keyword "zero" over different filterbanks.

TABLE I
PERFORMANCE COMPARISON

| Learning Method | Fix AFE Train DSCNN | Learn AFE and DSCNN | Learn AFE and DSCNN | |
|---|---|---|---|---|
| Loss Function | Cross-Entropy | Cross-Entropy | Co-design | |
| $\phi_I$ | 3.0 | 2.6 – 3.2 | 1.7 – 2.8 | |
| $\phi_C$ | 7.36 | 5.7 – 7.4 | 3.0 – 7.3 | |
| Q-factor | 4.78 | 1.9 – 4.8 | 1.7 – 4.5 | |
| Power (nW) | 96.4 | 97.2 | 88.0 | |
| Capacitor Area (mm$^2$) | 0.0653 | 0.0633 | 0.0569 | |
| Verification Method | Ideal Transfer Function | Ideal Transfer Function | Ideal Transfer Function | SPICE Simulation |
| Accuracy @ 5 dB SNR | 82.7% | 89.5% | 90.7% | 90.5% |
| Accuracy @ 20 dB SNR | 89.1% | 92.7% | 94.3% | 94.2% |

Fig. 12 shows the testing result of KWS accuracy over multiple SNR levels. The noise resilience is a notable feature achieved through SNR-aware training. Background noise segments are uniformly mixed into the training dataset with a probability of 0.8, spanning from 5 dB to 20 dB SNR. In contrast, noise-free training refers to training the model without mixing noise. To test the noise robustness of the trained model, the test dataset is iteratively evaluated by mixing environmental noise. Both SNR-aware training and noise-free training are applied to both the learnable AFE and fixed AFE, the results of which are illustrated in Fig. 12. The enhanced test accuracies at 5 dB SNR are 38.4% and 4.2% for SNR-aware training and learnable AFE, respectively.

The spike-count spectrograms, generated using the initial ideal transfer function, the learned ideal transfer function, the SPICE-simulated AC response, and the SPICE-simulated transient result, are visualized in Fig. 13. The spectrogram was produced using SPICE-simulated BPFs followed by behavioral models of the HWR and IAF stages. The learned AFE enhances feature contrast in the low channels and accentuates features in the middle channels. The difference between the spectrograms of the learned ideal AFE model and the SPICE-simulated model has minimal interference with the back-end classifier.

Fig. 14 shows the simulated BPF input-referred noise (IRN) spectral density before and after training using SPICE noise simulations. The 1$^{st}$, 8$^{th}$, and 16$^{th}$ channels results are shown in the figure as examples. Compared to the initial BPF IRN, the learned 1$^{st}$ channel and 16$^{th}$ channel BPFs have slightly higher IRN, while the 8$^{th}$ channel BPF has a nearly unchanged IRN. This is mainly due to the reduced gain in the 1$^{st}$ channel and 16$^{th}$ channel and the unchanged gain in the 8$^{th}$ channel. Nevertheless, through the co-optimization process with backend classifier, the reduced gain and increased IRN for these channels will not affect the overall classification accuracy. Furthermore, the integrated in-band input referred noise is 40.6 μV, 31.3 μV and 42.6 μV respectively. As the external noise from microphone is much higher than the circuit noise, BPF's intrinsic noise contribution was found to have only a minor effect on the overall performance. The transient simulation uses an input signal with amplitude of 5mV. This voltage level is based on the commercially available MEMS microphone performance to represent a normal conversation scenario. The BPF's intrinsic noise was included in the simulation, and its spectrogram has been compared in Fig. 13.

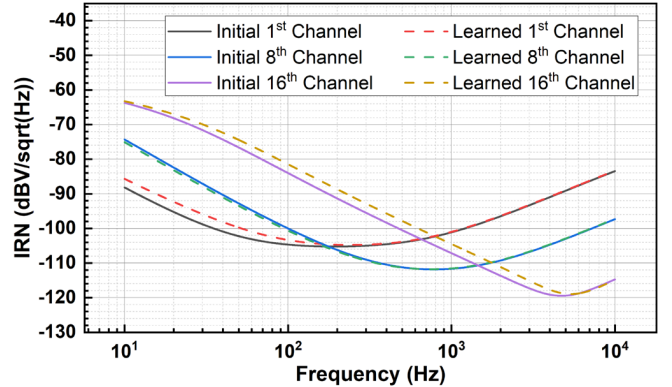

Fig. 14. Simulated BPF IRN spectral density for the 1$^{st}$, 8$^{th}$, and 16$^{th}$ channels before and after training.

*C. Benchmark and Discussion*

Table I shows the comparison between the fixed and learnable AFE on hardware resource utilization and KWS accuracy. Without a system-level loss function, the KWS accuracy can be improved by employing learnable AFE and SNR-aware training, obtaining 89.5% to 92.7% test accuracy under 5 dB to 20 dB SNR. However, it is challenging to co-optimize the hardware resource utilization simultaneously, since both power consumption and capacitor area are not considered during the training process. Through circuit-algorithm co-design and SPICE simulation, the presented architecture achieves notable reductions of 9.5% and 9.8% in a: Based on post-layout simulation.   b: Calculated based on information provided in paper.   c: SNR not reported.
d: Reported at 1% false alarm, -5 to 20 dB mixed SNR.   e: Without mixing noise into speech commands.

TABLE II
BENCHMARK TABLE

| | This Work | ISCAS 2023 [34] | JSSC 2023 [13] | JSSC 2022 [33] | ISSCC 2021 [15] | ICASSP 2023 [19] | ICLR 2020 [18] | VLSI 2019 [7] |
|---|---|---|---|---|---|---|---|---|
| **Feature Extraction (FEx)** | | | | | | | | |
| **FEx Type** | Analog | Digital | Analog | Analog | Analog | Digital | Digital | Digital |
| **No. of Channels** | 16 | 16 | 31 | 16 | 10 | 40 | 40 | 20 |
| **Learnable Filter** | Yes | No | No | No | No | Yes | Yes | No |
| **Process (Power Supply)** | 130 nm (0.6 V) | 65 nm (1.2 V) | 65 nm (0.4 V – 0.6 V) | 65 nm (0.5 V) | 65 nm (0.5 V) | NA | NA | 65 nm (0.4 V – 0.6 V) |
| **Power (nW)** | 113.7[a] | 3300[a] | 16.8 | 2990 | 109.0 | NA | NA | 8977 |
| **Area (mm$^2$)** | 0.77[a] | 0.02[a] | 0.228 | 1.6 | 0.5112 | NA | NA | 0.59[b] |
| **Keyword Spotting (KWS)** | | | | | | | | |
| **Benchmark Class (Keywords)** | GSCD 12 (10) | GSCD 12 (10) | GSCD 11 (10) | GSCD 12 (10) | GSCD NA (4) | GSCD 11 (10) | GSCD 35 (35) | GSCD 12 (10) |
| **Classifier** | DSCNN | LSTM | ResNet | GRU-FC | SNN | ResNet | CNN | LSTM |
| **Trainable Params. (Precision)** | 22k (INT16) | 22k (NA) | 238k (FP32) | 24k (INT14, INT8) | 69k (Binary) | 238k (FP32) | 4M (FP32) | 23k (INT8, INT4) |
| **Accuracy @ 5 dB SNR** | 90.4% | | | 9.8%[b] | | 87.4% | 78.0% | |
| **Accuracy @ 20 dB SNR** | 94.2% | 92.5%[c] | 91.5%[c] | 62.0% | 90.2%[d] | 94.9% | 85.0% | 90.9%[c] |
| **Accuracy @ base SNR [e]** | 94.5% | | | 86.0% | NA | 96.2% | 93.4% | |

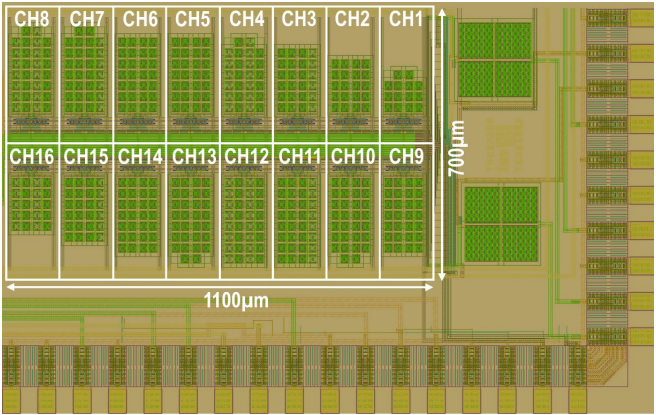

Fig. 15. Chip Layout in 130nm CMOS technology.

DSSF-BPF power and area consumption, respectively, while maintaining outstanding classification accuracy ranging from 90.5% to 94.2% under 5 dB to 20 dB SNR. The inclusion of $\phi_I$ and $\phi_C$ in the co-designed loss directly modifies the critical filter characteristics (transfer function including $f_c$, $Q$, and $A$), thereby optimizing the analog front-end preprocessing stage within the neural network pipeline. This joint optimization also serves as a regularization mechanism, preventing overfitting to the train dataset and avoiding overly complex circuit configurations. The effectiveness of this approach is demonstrated by a higher inference accuracy (94.3%) compared to the baseline cross-entropy-only training (92.7%).

Fig. 15 illustrates the chip layout of learned BPFs in 130 nm CMOS technology. The core active area is 1100µm × 700µm. In addition, the system performance is summarized and compared with other state-of-the-art designs in Table II, our design achieved the smallest classifier model size with outstanding classification accuracy. The power consumption is simulated under a 0.6 V power supply, and the FEx area is based on post-layout simulation. Without SNR-aware training, the inference accuracy dropped significantly at low SNR [33]. A digital IIR-based FEx is proposed in [34], which achieved comparable accuracy with smaller chip area. There is no fundamental limitation on scalability of the proposed co-design framework. However, as this design is based on time-domain analog bandpass filter bank for feature extraction, hence analog bandpass filter performances such as number of filters and filter bandwidth can be potential bottleneck for more complex tasks. This framework can be applied to low SNR scenarios such as 5 dB or below. However, covering a wider range of SNR will lead to peak classification accuracy degradation in higher SNR range. A promising future research direction is to enhance energy efficiency while maintaining classification accuracy across a wider range of SNRs by integrating approximate computing techniques into the classifier [8], [35].

## V. CONCLUSION

This paper presents a co-design approach integrating a learnable DSSF-BPF with a backend DSCNN for efficient audio classification. By incorporating adaptive scaling strategies, comprehensive circuit analysis, and a novel co-design optimization criterion, the framework achieves superior keyword spotting performance with efficient hardware resource utilization, requiring only 22k parameters. The DSSF-BPF structure addresses key challenges in analog feature extraction, such as power efficiency, area optimization, and system robustness, offering a flexible and energy-efficient solution. This approach not only ensures high classification accuracy and noise resilience but also facilitates fine-tuning of circuit parameters, making it ideal for low-power keyword spotting applications in IoT devices.

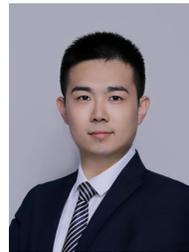

**Jinhai Hu** (Graduate Student Member, IEEE) received the B.Eng. degree in electrical and electronic engineering from the Nanyang Technological University (NTU), Singapore, in 2021. He is working toward his Ph.D. degree at NTU in collaboration with the Institute of Microelectronics (IME), Agency for Science, Technology and Research (A*STAR), Singapore. His current research interests include software-hardware co-design for biomedical signal analysis and human-machine interface.

J. Hu received the A*STAR Graduate Scholarship (AGS) in 2021 and the IEEE Circuit and System Society (CASS) Pre-Doctoral Grant in 2024.

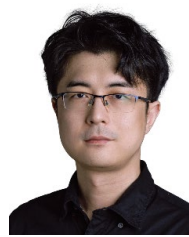

**Zhongyi Zhang** (Graduate Student Member, IEEE) received the B.Eng. degree in electrical engineering and its automation from Wuhan University, China, in 2018, and the M.Sc. degree in electronics from Nanyang Technological University (NTU), Singapore, in 2019. He is currently pursuing the Ph.D. degree in electrical and electronic engineering at NTU, Singapore.

From 2020 to 2024, he was a Research Associate with NTU. His current research interests include ultra-low-power circuits, analog front-end design, acoustic feature extraction, and bandpass filters.

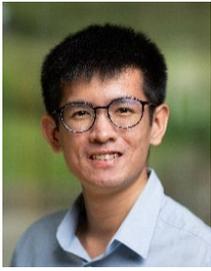

**Cong Sheng Leow** (Graduate Student Member, IEEE) received the B.Eng. degree in electrical and electronic engineering from the Nanyang Technological University (NTU), Singapore, in 2022. He is currently pursuing the Ph.D. degree in electrical and computer engineering with the University of Michigan, Ann Arbor, MI, USA.

From 2022 to 2023, he was a Research Engineer with the Institute of Microelectronics (IME), Agency for Science, Technology, and Research (A*STAR), Singapore, His current research interests include mixed-signal systems, emerging computing paradigms, and biomedical applications.

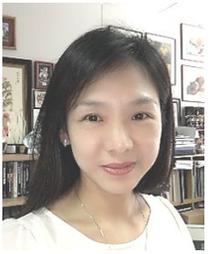

**Wang Ling Goh** (Senior Member, IEEE) received both her Bachelor of Engineering Degree in Electrical and Electronic Engineering and Doctor of Philosophy in Microelectronics from the Department of Electrical and Electronic Engineering at the Queen's University of Belfast in United Kingdom in 1990 and 1995, respectively.

Dr Goh joined the School of Electrical and Electronic Engineering (EEE) as a lecturer and became an Associate Professor in 2004. Prior to her current appointment as Associate Dean (Academic) at the Graduate College, she had held many academic positions such as Deputy Director (Undergraduate) of the Renaissance Engineering Programme (Jun 2019 – Jul 2021), Coordinator for the Final Year Projects at School of EEE (Sep 2018 – Jul 2020), Programme Coordinator of B.Eng. (EEE) (Jun 2014 – May 2017), Member of NTU Teaching Council (Oct 2012 – Jun 2013), Associate Dean (Outreach & External Relations) at the College of Engineering (Jan 2010 – Dec 2012), Assistant Chair of Students (Jul 2008 – Dec 2009) and Assistant Head of Division at School of EEE (Sep 2006 – Jun 2008).

Dr Goh participates actively as General Chair or Advisory/Technical Committee Member in various international conferences. She was the General Conference Chair of the 2016 International Symposium on Integrated Circuits (ISIC 2016), held in Singapore, from December 12-14 in 2016. Dr Goh's research interests include digital/mixed-signal.

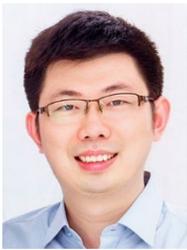

**Yuan Gao** (Member, IEEE) received the B.E and M.E degrees in electrical engineering from the Huazhong University of Science and Technology, Wuhan, China, in 2000 and 2002, respectively, and the Ph.D. degree in electrical engineering from the National University of Singapore, Singapore, in 2008.

Since 2007, he has been with the Institute of Microelectronics (IME), Agency for Science, Technology and Research (A*STAR), Singapore. He is currently a principal investigator and principal scientist in the Integrated Circuit Design and Systems (ICDS) Department, where he is leading the next generation intelligent sensor interface IC development. He has authored or coauthored 3 book chapters, more than 120 peer-reviewed international journal and conference papers and has more than 10 US patents granted or filed. He has co-supervised 8 PhD students and he is an accredited A*STAR PhD Scholar supervisor. He received A*STAR Graduate Academy Star Mentor Award in 2023 and IEEE Solid State Circuit Society Outstanding Reviewer Award in 2023. His primary research areas include energy efficient analog and mixed-signal IC design in the emerging areas such as intelligent sensor interface, AI hardware, biomedical microsystem and energy harvesting.

Dr. Gao was TPC member of the IEEE International Solid-State Circuits Conference (ISSCC) between 2015 – 2020 and served as Associate Editor of the IEEE TRANSACTIONS ON CIRCUITS AND SYSTEMS−I: REGULAR PAPERS between 2020 – 2022. Currently he is an Associate Editor of the IEEE TRANSACTIONS ON BIOMEDICAL CIRCUITS AND SYSTEMS.